\documentclass[12pt,preprint]{aastex}
\usepackage{subfigure}
\usepackage{natbib}
\usepackage{epsfig,amsmath,amssymb,amsfonts,mathrsfs,latexsym,graphicx,lscape}
\usepackage{hyperref}
\usepackage{times}
\usepackage{epsfig}
\usepackage{rotating}
\usepackage{sidecap}
\usepackage{longtable}

\shorttitle{Cross-calibration X-ray Detectors with Scaling Relations of Galaxy Clusters}
\shortauthors{zhaohh et al.}
\begin{document}

\title{Unbiased Correction Relations for Galaxy Cluster properties Derived from  {\it Chandra} and {\it XMM-Newton}
}

\author{Hai-Hui Zhao,  Cheng-Kui Li, Yong Chen, Shu-Mei Jia, Li-Ming Song}

\altaffiltext {} {Particle Astrophysics Division,
Institute of High Energy Physics,
 Chinese Academy of Sciences,
 Beijing 100049; zhaohh@ihep.ac.cn}

\begin{abstract}

We use a sample of 62 clusters of galaxies to investigate the discrepancies of gas
temperature and total mass within $r_{500}$ between {\it XMM-Newton} and {\it Chandra} data.
Comparisons of the properties show that£º (1)
Both the de-projected and projected  temperatures determined by {\it Chandra}  are higher than those of {\it XMM-Newton}
and there is a good linear relation for the de-projected temperature:
$ T_{Chandra}$=1.25$\times  T_{XMM}$-0.13.
(2) The {\it Chandra} mass is much higher than {\it XMM-Newton} mass with a bias of 0.15 and our mass relation
is: $\log_{10} M_{Chandra}$=1.02 $\times \log_{10} M_{XMM}$+0.15.
To explore the reasons for the discrepancy in mass, we recalculate the {\it Chandra} mass (expressed as $M_{Ch}^{mo/d}$) by modifying
its temperature with the de-projected temperature relation.
The results show that $M_{Ch}^{mo/d}$ is more close to the {\it XMM-Newton} mass  with the bias reducing  to 0.02.
Moreover, $M_{Ch}^{mo/d}$ are corrected with the $r_{500}$ measured by {\it XMM-Newton} and the intrinsic scatter is
significantly improved with the value reducing from  0.20 to 0.12.
These mean that the temperature bias may be the main factor causing the mass bias.
At last, we find that  $M_{Ch}^{mo/d}$ is consistent with the corresponding {\it XMM-Newton} mass  derived directly from
our mass relation at a given {\it Chandra} mass.
Thus, the de-projected temperature and  mass relations can provide  unbiased corrections  for galaxy cluster properties
derived from {\it Chandra} and {\it XMM-Newton}.

\end{abstract}


\section{Introduction}
Galaxy clusters are the most massive gravitationally bound objects in the Universe, and they
can provide crucial information for studies of large scale-structure (Bahcall 1988;
Zhang et al. 2006; Willis et al. 2013) and tracing cosmic evolution
(Allen et al. 2011).
The cluster mass is probably the most interesting global parameter for characterizing a galaxy cluster (B\"ohringer et al. 2004).
The cluster mass function, which is  sensitive to cosmological parameters, can give observational
constraints to cosmology (Vikhlinin et al. 2009,  Tinker et al. 2012).
Precise mass estimate strongly depends on the measurement of gas temperature in the cluster.
By accurately measuring temperatures and masses of  galaxy clusters in a large sample, one can  calibrate
the mass-temperature relation, which is widely used to improve the accuracy of the cosmological parameters
determination (Nevalainen et al. 2010).

Precise properties of galaxy cluster can be derived from {\it XMM-Newton}
and {\it Chandra}, which have high spatial resolution and large
field of view. However, there are discrepancies in the measurements of gas temperature and total mass between the two
instruments. Typically, the {\it Chandra} temperature  is 5\%-15\% higher than the value of {\it XMM-Newton}
(Kotov \& Vikhlinin 2005; Vikhlinin et al. 2005; Snowden et al. 2008; Reese et al. 2010).
The total mass  within $r_{2500}$ derived from {\it Chandra} was
roughly 15\% higher  than that from {\it XMM-Newton} (Mahdavi et al. 2013).

Many works have attempted to modify the systematic differences in the cluster's temperature or mass between these two instruments.
By multiplying the effective area of {\it Chandra}/ACIS with the
corresponding splines of the stacked residuals, Schellenberger et al. (2014) changed the energy dependence
of effective area and found that the temperatures between {\it Chandra}/ACIS and {\it XMM-Newton}/pn were  consistent.
Li et al. (2012) tried to fit {\it Chandra} spectra with  {\it XMM-Newton} temperatures and presented that
the modified  {\it Chandra} mass of Abell 1835 was consistent with the {\it XMM-Newton} mass.
These works focused on looking for the reasons of discrepancy in temperature or mass,
but they didn't  give an  correction  relation which can be used directly in combining
{\it Chandra} and {\it XMM-Newton} data to build a large sample.

In this work, we use a sample of 62 clusters of galaxies
to study the discrepancies of temperature and mass derived from  {\it Chandra} and {\it XMM-Newton},
 aiming to find good correction methods for the  discrepancies of properties.
This paper is organized as follows.
Section 2 shows the reduction procedures for {\it Chandra} and {\it XMM-Newton} data and the methods to obtain
 temperature profile and total mass.
The relations of  temperature and mass  between the two instruments are listed in section 3.
In Section 4, we attempt to illustrate the temperature bias is the main factor causing the mass discrepancy. We
prove our mass relation between the two instruments is  robust to correct the {\it Chandra}  mass, which can
bring the masses obtained with  {\it Chandra} and {\it XMM-Newton} into consistency in Section 5.
We draw our conclusions in Section 6.
Throughout this paper, the selected energy band is 0.5-7.0 keV and a flat $\rm \Lambda$CDM cosmological model is used with
$\Omega_{\rm M}$ = 0.3, $\Omega_{\rm \Lambda}$ = 0.7 and $H_{0}$ = 70 km s$^{-1}$ Mpc$^{-1}$.

\section{Sample Selection and Data  Analysis}

Using a flux-limited ($f$ $\geq$ 1.0 $\times$ $10^{-11}$  {$\rm erg$ $\rm s^{-1}$ $\rm cm^{2}$})
method, we have built a large cluster sample from the
RASS (Grandi et al. 1999), HIFLUGCS (Thomas et al. 2002), REFLEX (B\"ohringer et al. 2004),
NORAS (B\"ohringer et al. 2000), XBACs (Ebeling et al. 1996) and BCS (Ebeling et al. 1998)
catalogs. In this sample, there are  72 clusters observed  by  both {\it Chandra} and {\it XMM-Newton}.
We analyze all of the clusters to find the overall biases of cluster properties between the two instruments.
Data reduction procedures for {\it Chandra} and {\it XMM-Newton} are as follows.

The {\it XMM-Newton} data are processed with Science Analysis System (SAS) 12.6.0.
In this paper, we consider the pn/EPIC data which have larger effective area. The
observations are taken in Extended  Full Frame mode or Full Frame mode.
The events with FLAG = 0, PATTERN $\leq$ 4 are used, the read out of time
and the vignetting effects are also corrected.
Since the X-ray flux of the cluster should be stable during the observation period, we discard all the intervals
with prominent flares and then select only those intervals with count rates within 3$\sigma$ of  the residual average
value.
 After the  removal of  the prominent background flares and point sources,
the observation of $'$Lockman Hole$'$ (observation ID: 0147511801,hereafter LH) is used to subtract the background.
Considering the background difference between the LH and the
 source, we use the local background to monitor the residual background.
The data reduction procedures for {\it XMM-Newton} can be referred
to  Zhao et al. (2013).

The {\it Chandra} data are performed by  CIAO 4.3 and CALDB 4.4.0.
We analyze the {\it Chandra} data following  the method  discussed in Li et al. (2012).
The tool LC\_CLEAN in CIAO  is used to scan the  light curve of data for flares,
and the Good Time Intervals (GTIs) are selected.  The prominent background flares are removed as
in {\it XMM-Newton}. We extract background from the standard set of CTI-corrected ACIS blank sky  in
the  {\it Chandra} CALDB (Markevitch et al. 2003) and the process is the same as {\it XMM-Newton}.

For both  {\it XMM-Newton} and {\it Chandra} data,
a double-background subtraction method is applied  to correct  the Particle background and
Cosmic X-ray Background as used in Jia et al. (2004, 2006).
Assuming spherical symmetry,  the spectra are extracted from annular regions centered on the X-ray emission peak.
The criterion of $\sim$ 2000 net counts in 2-7 keV band per bin is used to determine the width of each
ring (Zhang et al. 2006, 2007). The minimum width of the rings are set at 0.5$'$ or 0.25$'$
for {\it XMM-Newton}/EPIC or {\it Chandra}/ACIS, respectively. It is wide enough for us to ignore
the Point Spread Function (PSF) effect.
The de-projected spectrum of each shell is derived by  subtracting
all the contributions from the outer regions (see Chen et al. 2003, 2007 and Jia et al. 2004,
2006 for detailed calculation).

%----------------------------------------------------------------------------------------------------------

\subsection{De-projected Temperature Profile and Total Mass}

The spectral analysis is carried out using XSPEC version 12.8.1.
The plasma emission model Mekal (Mewe et al. 1985) and
WABS model (Morrisson \& McCammon 1983) are used to fit the de-projected spectra and then  the de-projected temperature,
 metallicity  and normalizing constant $ norm$ in each ring can be obtained.
We  fit the radial de-projected temperature profile
by the following equation (Xue et al. 2004):
\begin{equation}
{T(\rm r)} = {T_{\rm 0}}+\frac{A}{r/r_{\rm 0}}\exp(-\frac{(\ln r-\ln r_{\rm 0})^{2}}{\omega}),
\end{equation}
where $T_{\rm 0}$, $A$, $r_{\rm 0}$, and $\omega$ are free parameters.

For the electron density profile, we divide the cluster into several annular regions (10-30 regions,
depending on count rate of the cluster) centered on the emission peak. Then, the de-projected photon counts
in each shell are obtained. Since the de-projected temperature and abundance profiles are known, we  can
simulate the spectrum of each shell  in XSPEC. By fitting the simulative spectra,
 $ norm$ of each shell are determined, which can provide the corresponding
electron density $n_{\rm e}$ (Jia et al. 2006).
A double-$\beta$ model is adopted to fit the de-projected electron density profile (Chen et al. 2003):
\begin{equation}
n_{\rm e}(r) = n_{\rm 01}[1+(\frac{r}{r_{c1}})^{2}]^{-\frac{3}{2}\beta_{1}}+ n_{\rm 02}[1+(\frac{r}{r_{c2}})^{2}]^{-\frac{3}{2}\beta_{2}},
\end{equation}
where $n_{\rm 01}$ and $n_{\rm 02}$ are electron number density parameters, $\beta_{1}$ and $\beta_{2}$ are
the slope parameters, and $r_{\rm c1}$ and $r_{\rm c2}$ are the core radii
of the inner and outer components, respectively.
Then,  with the assumptions of hydrostatic equilibrium and spherical symmetry,
the gravitational mass of cluster within radius $r$ can be determined as (Fabricant et al. 1980):
\begin{equation}
M( < r ) = -\frac{k_{\rm B}Tr^{2}}{G\mu m_{\rm p}}[\frac{d(\ln n_{\rm e})}{dr} + \frac{d(\ln T)}{dr}].
\end{equation}
where $\mu$ is  the mean molecular weight of gas and the value is assumed to 0.62.
$k_{\rm B}$ is the Boltzman constant, $m_{\rm p}$ is  the proton mass, and $G$ is  the gravitational
constant. Hereafter,  $M_{Chandra}$ and $M_{XMM}$ represent the original {\it Chandra}
and {\it XMM-Newton}  mass within $r_{\rm 500}$, in
which the mean gravitational mass density is equal to 500 times the
critical density at the cluster redshift.
The primary parameters of all 62 galaxy clusters are given in Table 1.

\subsection{Statistical Analysis}

In the following, we investigate the temperature and mass  relationships between {\it Chandra} and {\it XMM-Newton} data.
The commonly used BCES Bisector method (Akritas \& Bershady 1996) is used to fit the relations.
We perform the relations of temperature and mass in the form:

\begin{equation}
Y =B \times X + A ,
\end{equation}
where $A$ and $B$ are the two free parameters to be estimated, $X$ and $Y$ represent
the temperatures ($T_i$) or the logarithmic values of  masses ($\log_{10} M_i$)  derived
from {\it XMM-Newton} and {\it Chandra}. The intrinsic scatter around the best-fit relation
is  calculated as (Morandi et al. 2007; Pratt et al. 2009):
\begin{equation}
S = \left[ \sum_{i}w_i
\left(\left( Y_{i} -B \times X_{i} -A  \right)^2
-\sigma_{Y_{i}}^2\right) / (N-2) \right]^{1/2}\ ,
\end{equation}

\noindent where

\begin{equation}
w_i = \frac{1/\sigma_i^2}{(1/N) \sum_{i=1}^N 1/\sigma_i^2}\ \  {\rm ,}\ \ \sigma_i^2 = \sigma^2_{Y_i} + B^2 \sigma^2_{X_i},
\end{equation}

\noindent
$N$ is the total number of data, $\sigma_{Y_i}$ and $\sigma_{X_i}$ are the statistical errors of the measurements $Y_i$ and $X_i$, respectively.

In order to evaluate the systematic deviation, we define the bias of  temperature or mass measurement
as  the average vertical distance  between the best-fit line and the line of $Y$=$X$ ({\it bias = $\frac{1}{N}$$\sum$ (Y$_i$- X$_i$})).
 The results of fits for all the relations are given in Table 3.

%--------------------------------------------------------------------------------------------------------
\section{The Relations of Properties between {\it Chandra} and {\it XMM-Newton}}

\subsection{The Relations of Temperatures Determined by  {\it Chandra} and {\it XMM-Newton}}
\begin{figure}
\centerline{\hbox{\psfig{figure=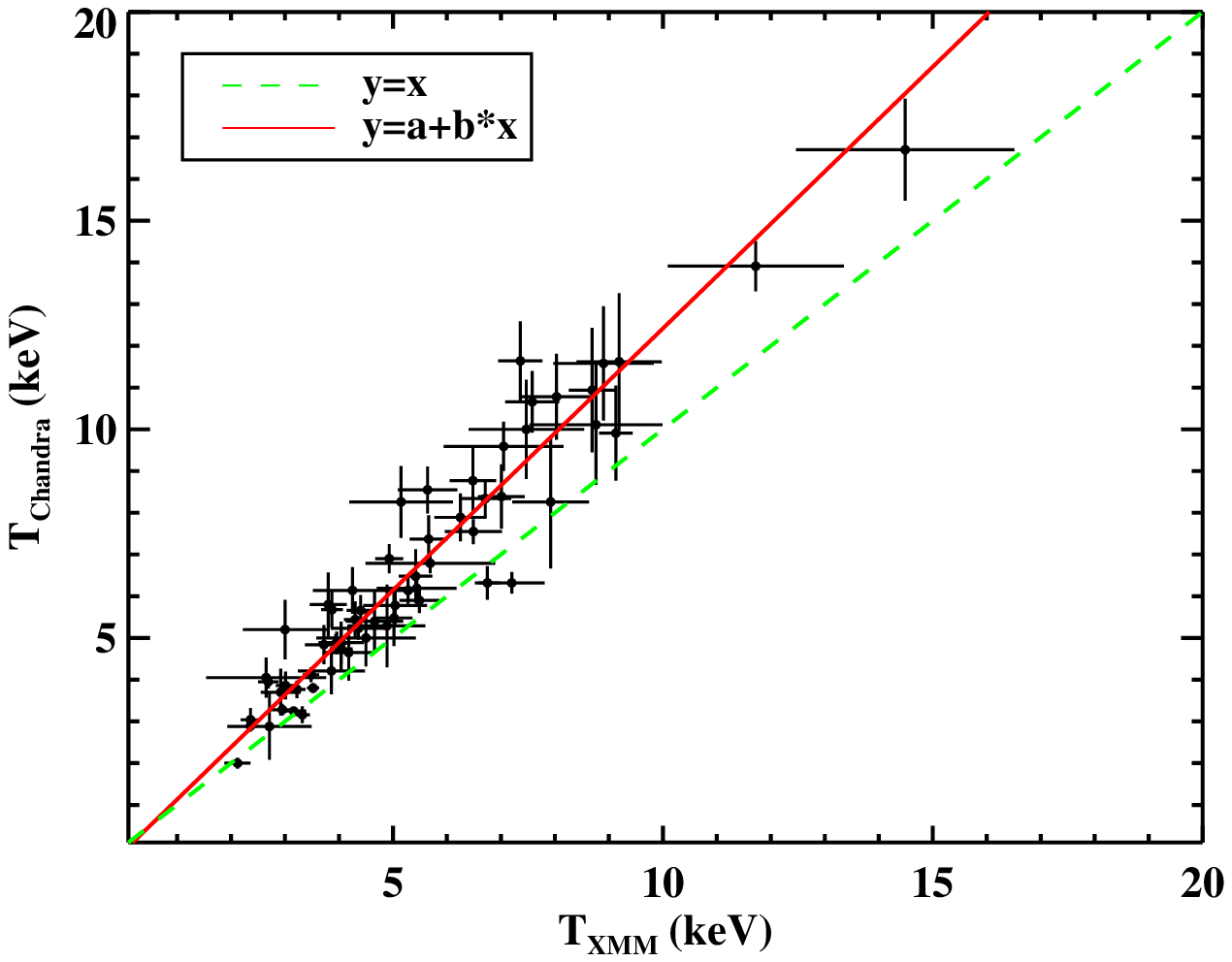,angle=0,scale=0.55}
\psfig{figure=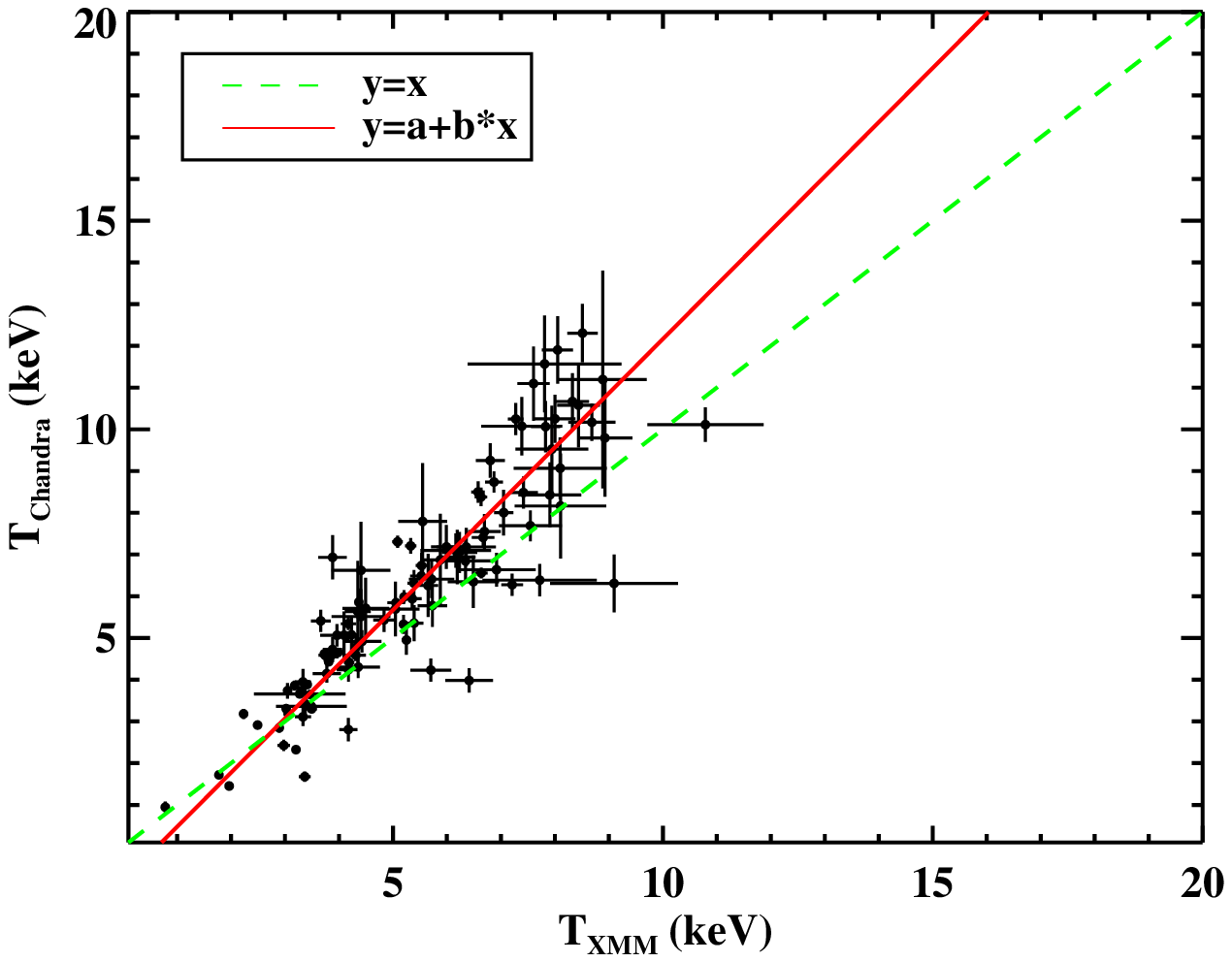,angle=0,scale=0.55}}}
\caption[Comparisons of temperatures between {\it Chandra} and {\it XMM-Newton}] {Left panel: comparison
of global temperatures determined by {\it Chandra} and  {\it XMM-Newton}.
The global temperatures are the volume averages of the de-projected temperatures which are
limited to the  ranges of 0.15-0.5 $r_{500}$ and 0.2-0.5 $r_{500}$ for $T_{Chandra}$
and $T_{XMM}$, respectively; Right panel:  comparison of {\it Chandra} and
{\it XMM-Newton} projected temperatures.
For each cluster, we derive
two projected temperatures within fixed rings (the radii are 1.0$'$-2.5$'$ and 2.5$'$-4.0$'$).
The red lines show the best-fit temperature relations, while the green dashed lines show the relations of $Y$=$X$.}
\label{zhaohh:fig1}
\end{figure}

For each cluster, the de-projected temperatures are obtained in several rings.
To avoid the effects of cool cores and keep the qualities of spectral data, we derive the global  temperatures by the volume
average of the de-projected temperatures, which is in the radii of  0.15-0.5$r_{500}$ (Vikhlinin et al. 2009)
and 0.2-0.5$r_{500}$ (Zhang et al. 2008) for {\it Chandra}  and {\it XMM-Newton} data , respectively.
Due to the effects of PSF, a larger inner boundary of 0.2$r_{500}$ is set for {\it XMM-Newton} data.
The maximum radii ($R_{max}$), out to which the temperature profiles can reach
for each cluster as a fraction of $r_{500}$, are calculated. We list the global temperatures and $R_{max}$ in Table 1.
The extended temperature and electron density profiles may introduce some uncertainties
to the cluster mass. Ten clusters (e.g., 2A0335, A1060, A262),
whose  $r_{500}$ are much larger than the field of view (the $R_{max}$ is smaller
than 0.5$r_{500}$), are not considered in the following analysis.
Comparing the global  temperatures between the two instruments,
the de-projected temperature derived from {\it Chandra} is higher than that of {\it XMM-Newton} by about 1.24 keV
as shown in the left panel of  Fig. \ref{zhaohh:fig1}.
There is a good linear relation for the de-projected temperatures measured by the  two instruments and the best-fit relation
is $T_{Chandra}$=1.25$\times T_{XMM}$-0.13 with the  intrinsic scatter of 0.50.

In order to find the discrepancy of temperatures between {\it Chandra} and {\it XMM-Newton} directly,
we extra the projected spectra  within two  fixed rings
(the radii are 1.0$'$-2.5$'$ and 2.5$'$-4.0$'$) and get the projected temperatures
for each cluster. The comparison of projected temperatures between the two instruments
is shown in the right panel of  Fig. \ref{zhaohh:fig1}.
There still exist a linear relation between the two temperatures and the relation is $T_{Chandra}$=1.30$\times T_{XMM}$-0.83  with
the  intrinsic scatter of 0.57.
The projected temperature obtained with {\it Chandra} is higher than that of {\it XMM-Newton} by about 0.79 keV, which is smaller
than the bias of de-projected temperature.
Our discrepancy (11\%) of the projected temperatures is consistent with the value of 10-15\% in Nevalainen et al. (2010).

%\clearpage

\subsection{The Relation of Masses Derived from {\it Chandra} and {\it XMM-Newton}}

\begin{figure}
\begin{center}
\includegraphics[angle=0,scale=0.55]{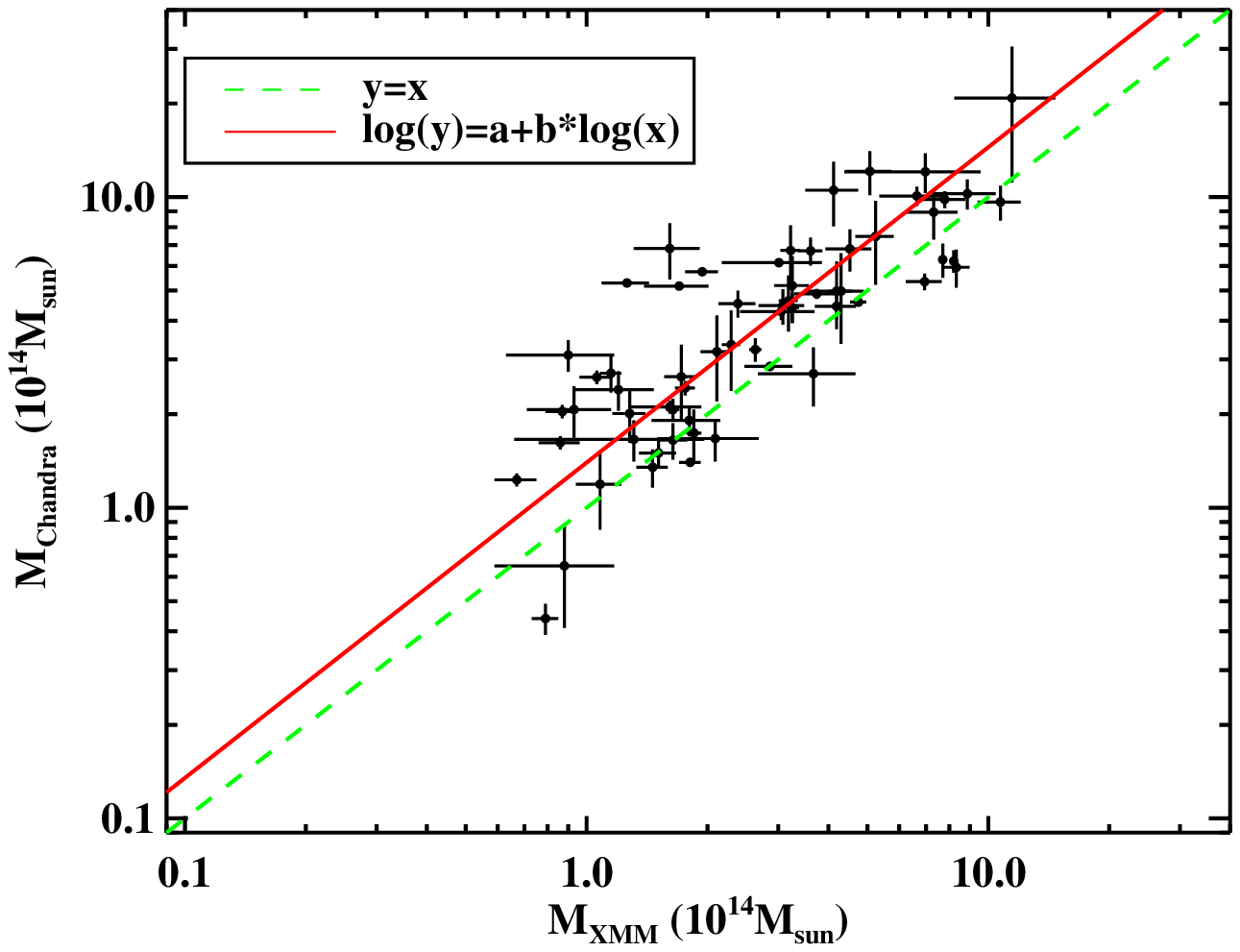}
\caption{Comparison of original de-projected  masses obtained with  {\it Chandra} and  {\it XMM-Newton}.}
\label{zhaohh:fig2}
%\end{flushright}
\end{center}
\end{figure}

Using the  radial de-projected temperature and intra-cluster medium density distributions, we obtain
the total mass within $r_{500}$ for both  {\it Chandra} and {\it XMM-Newton}. Comparison of the total masses
between the two instruments is shown in  Fig. \ref{zhaohh:fig2}.
There is a logarithmic relationship for the masses  between the two
instruments: $\log_{10}M_{Chandra}$=1.02 $\times \log_{10}M_{XMM}$+0.15, and the
intrinsic scatter around the this  relation is 0.19.
Fixing the slope of our mass relation to 1.0, we find that the mass determined by {\it Chandra}
is higher about 36\% that that of  {\it XMM-Newton}.
Using  our definition of bias in Section 2.2, the bias between the two masses is 0.15.
We also obtain the total mass within $r_{2500}$  for  both {\it Chandra} and {\it XMM-Newton}
 and find a difference of about 20\% which is larger than  the result of 15\% for 19 clusters in  Mahdavi et al. (2013).

%\clearpage

\section{The Effects of Temperature on Mass Discrepancy}

\begin{figure}
\centerline{\hbox{\psfig{figure=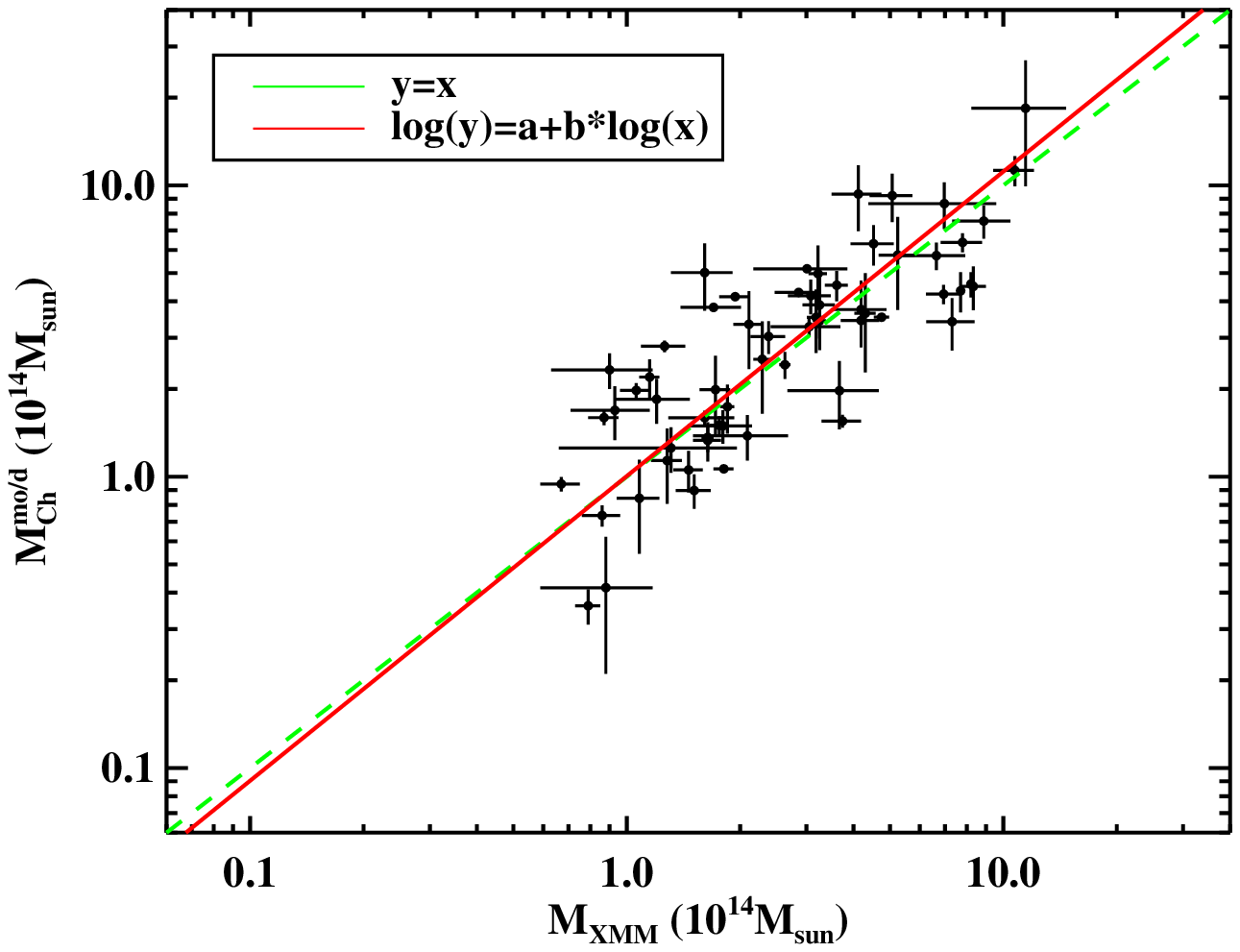,angle=0,scale=0.55}
\psfig{figure=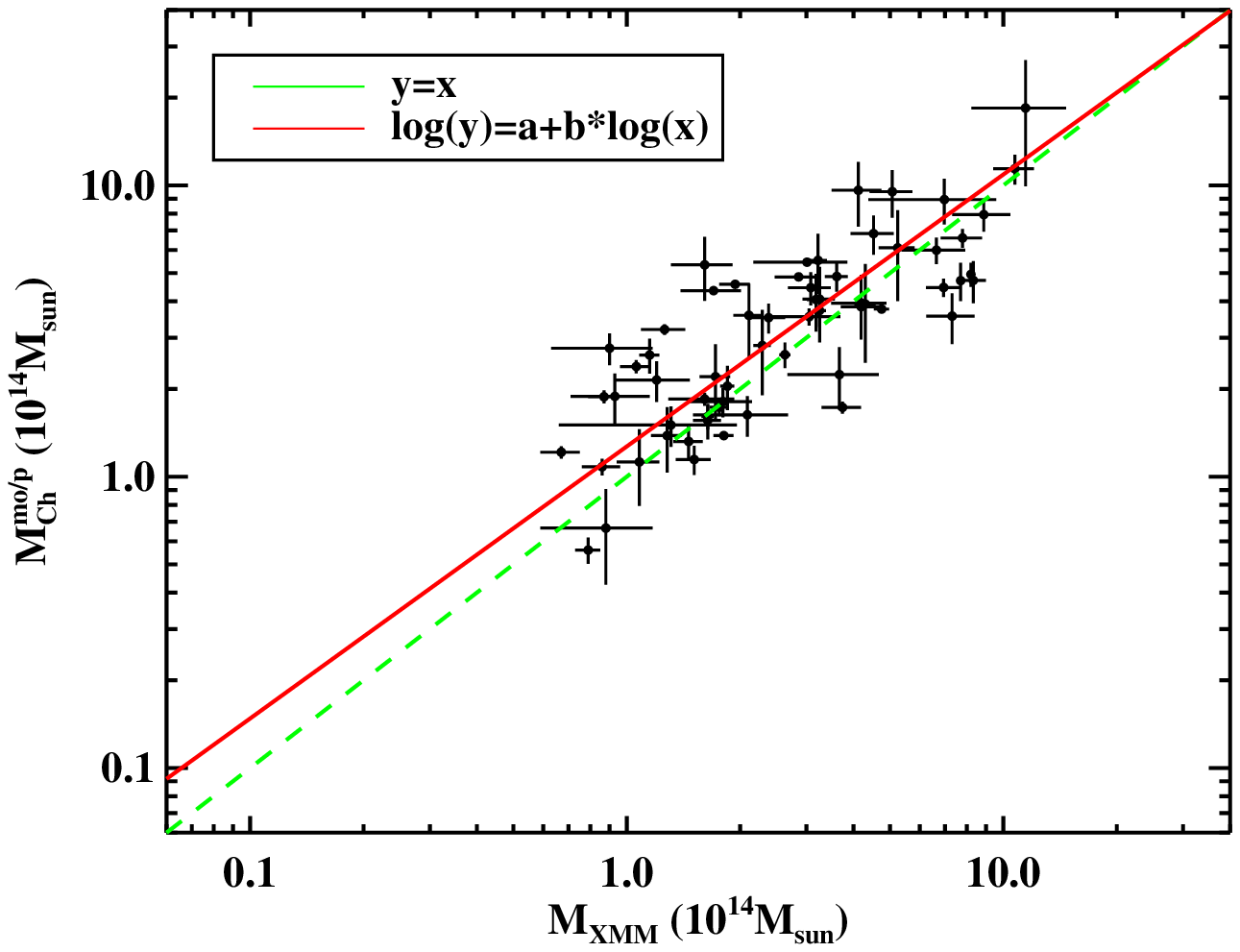,angle=0,scale=0.55}}}
\caption[Comparisons of corrected {\it Chandara} masses and {\it XMM-Newton} masses.] {Comparison of
{\it XMM-Newton}  and  corrected {\it Chandra} masses. Left panel: the {\it Chandra} masses are
modified by the relation of de-projected temperatures; Right panel: the {\it Chandra} masses are modified by the relation of
projected temperatures.}
\label{zhaohh:fig3}
\end{figure}

The  results above indicate that both the temperature and mass derived from {\it Chandra} are higher than the
values obtained with {\it XMM-Newton}. It is interesting to test whether the discrepancy in mass is caused by temperature.
Many recent works reveal that the properties derived from {\it XMM-Newton} are more reliable (B\"ohringer et al. 2004; Zhang et
al. 2010; Li et al. 2012),
and we try to modify the {\it Chandra} mass with the relations of  temperatures  in this paper.
Using the relations of temperatures between the two instruments, we correct the {\it Chandra} temperature
of each ring and the  corrected temperature profiles for {\it Chandra} data are obtained. Based on the modified {\it Chandra}
temperature, we recover the electron density distribution.
Then, we get the amended {\it Chandra} masses ($M_{Ch}^{mo/d}$ and $M_{Ch}^{mo/p}$ for the {\it Chandra} mass modified
by the de-projected and projected temperature relations, respectively).
We compare the modified {\it Chandra} masses with the $M_{XMM}$ as shown in Fig. \ref{zhaohh:fig3}.
Our results show  that both the modified {\it Chandra} masses are more close to the $M_{XMM}$, the bias
 between $M_{XMM}$ and $M_{Ch}^{mo/d}$ is 0.02 while the bias  between $M_{XMM}$ and $M_{Ch}^{mo/p}$ is 0.08.
Compared with the bias of 0.15 in Section 3.2,  the  mass bias is almost resolved by the temperature correction, this
means the temperature may be the main factor causing the mass bias.

%\clearpage

\begin{figure}
\centerline{\hbox{\psfig{figure=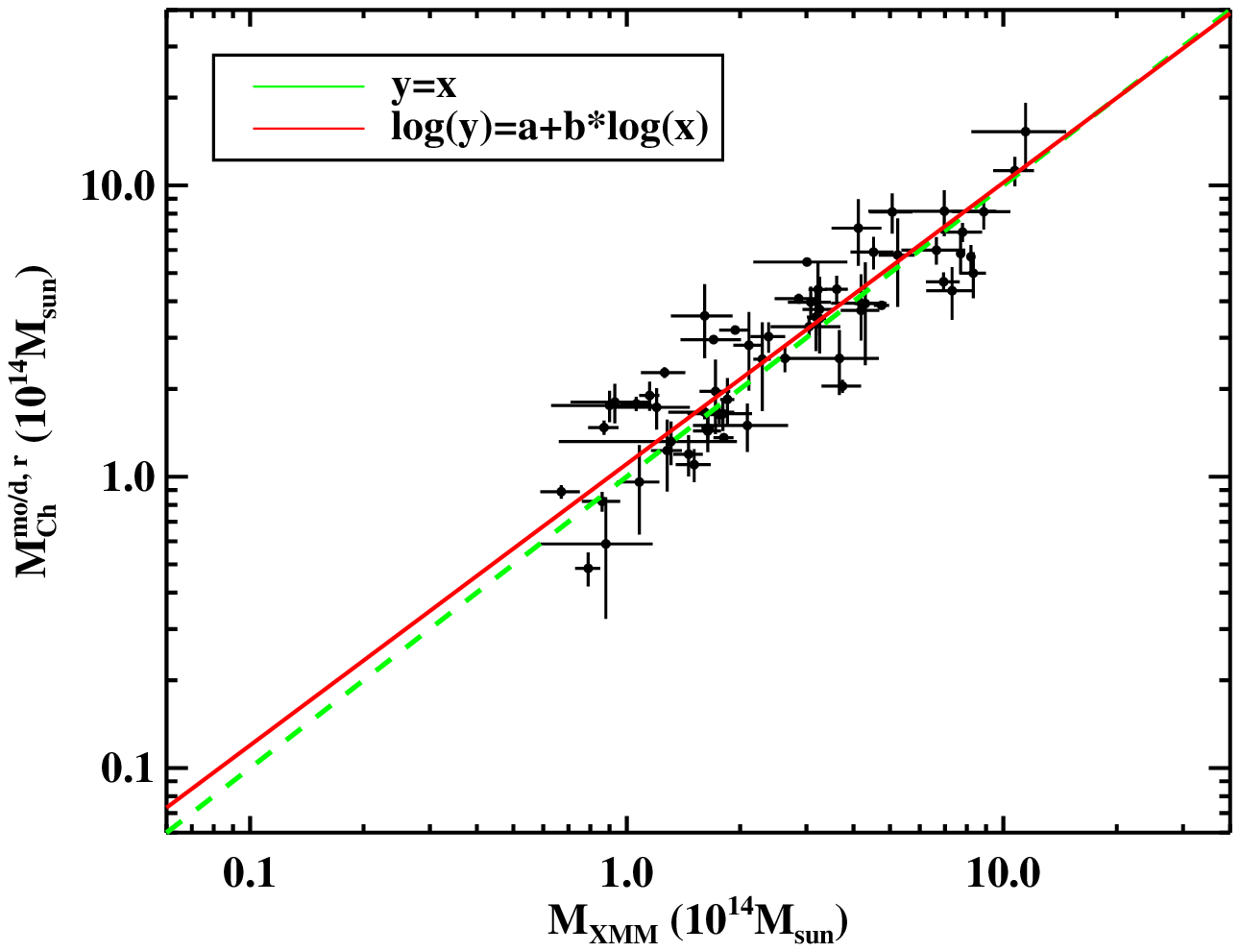,angle=0,scale=0.55}
\psfig{figure=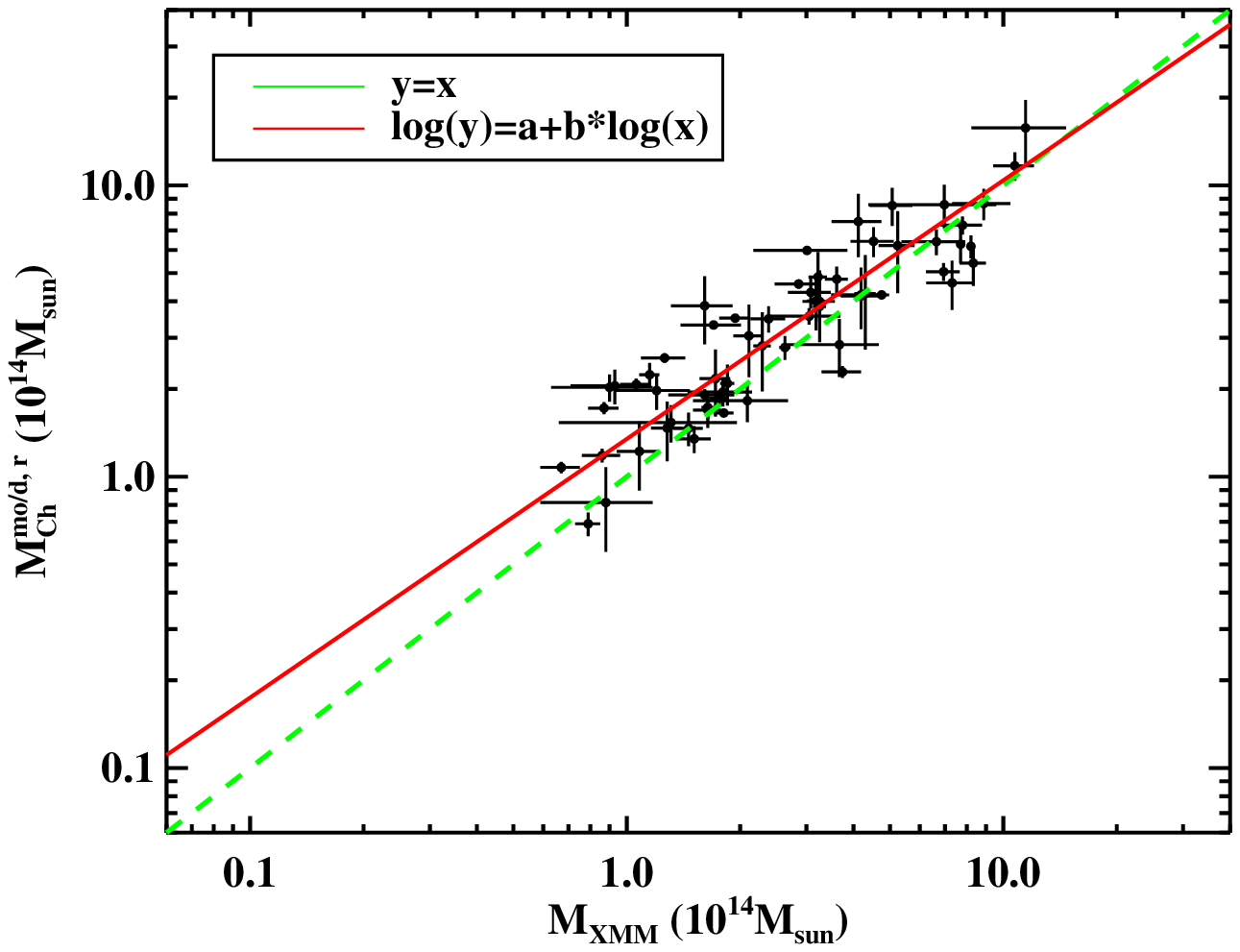,angle=0,scale=0.55}}}
\caption[Comparisons of modified {\it Chandara} masses and {\it XMM-Newton} masses.] {
Same as Fig. 3, excepting that  the corrected {\it Chandra} masses are integrated  to the $r_{500}$ measured by {\it XMM-Newton}.}
\label{zhaohh:fig4}
\end{figure}

After correcting the {\it Chandra} mass, the  bias of mass between the two instruments is indeed smaller than before,
 but the intrinsic scatters are not improved as shown in Table 3.
In the mass comparisons above, the  masses derived from different instruments are integrated within their own
$r_{500}$, and different $r_{500}$  may also bring bias to the mass determination. To reduce such discrepancy, we recalculate
the {\it Chandra} mass with the   $r_{500}$ measured by {\it XMM-Newton}
(the modified {\it Chandra} masses are expressed  as $M_{Ch}^{mo/d,r}$ and $M_{Ch}^{mo/p,r}$).
The comparisons of $M_{XMM}$ with $M_{Ch}^{mo/d,r}$ and $M_{Ch}^{mo/p,r}$ are shown in Fig. \ref{zhaohh:fig4} and the results are listed in Table 3.
There are obvious improvements in the intrinsic scatters: the intrinsic scatter is  reduced from 0.20 to 0.12 for
the $M_{XMM}$-$M_{Ch}^{mo/d,r}$ relation; and the value change from 0.18 to 0.11 for the $M_{XMM}$-$M_{Ch}^{mo/p,r}$ relation.
The  $M_{Ch}^{mo/d,r}$ is still in agreement with $M_{XMM}$ with the bias of $0.03 \pm 0.01$,
while the $M_{Ch}^{mo/p,r}$ is higher than  $M_{XMM}$ with the bias of $0.08 \pm 0.01$
as shown in Fig. \ref{zhaohh:fig4}.
This reveals that the corrected method using the de-projected
temperature relation is more effective.  Thus,  after correcting the temperature and $r_{500}$, the
intrinsic scatter is smaller than before and the $M_{Ch}^{mo/d,r}$ is consistent with $M_{XMM}$.

\section{Comparison of Methods for Correcting {\it Chandra} Mass}

\begin{figure}
\centerline{\hbox{\psfig{figure=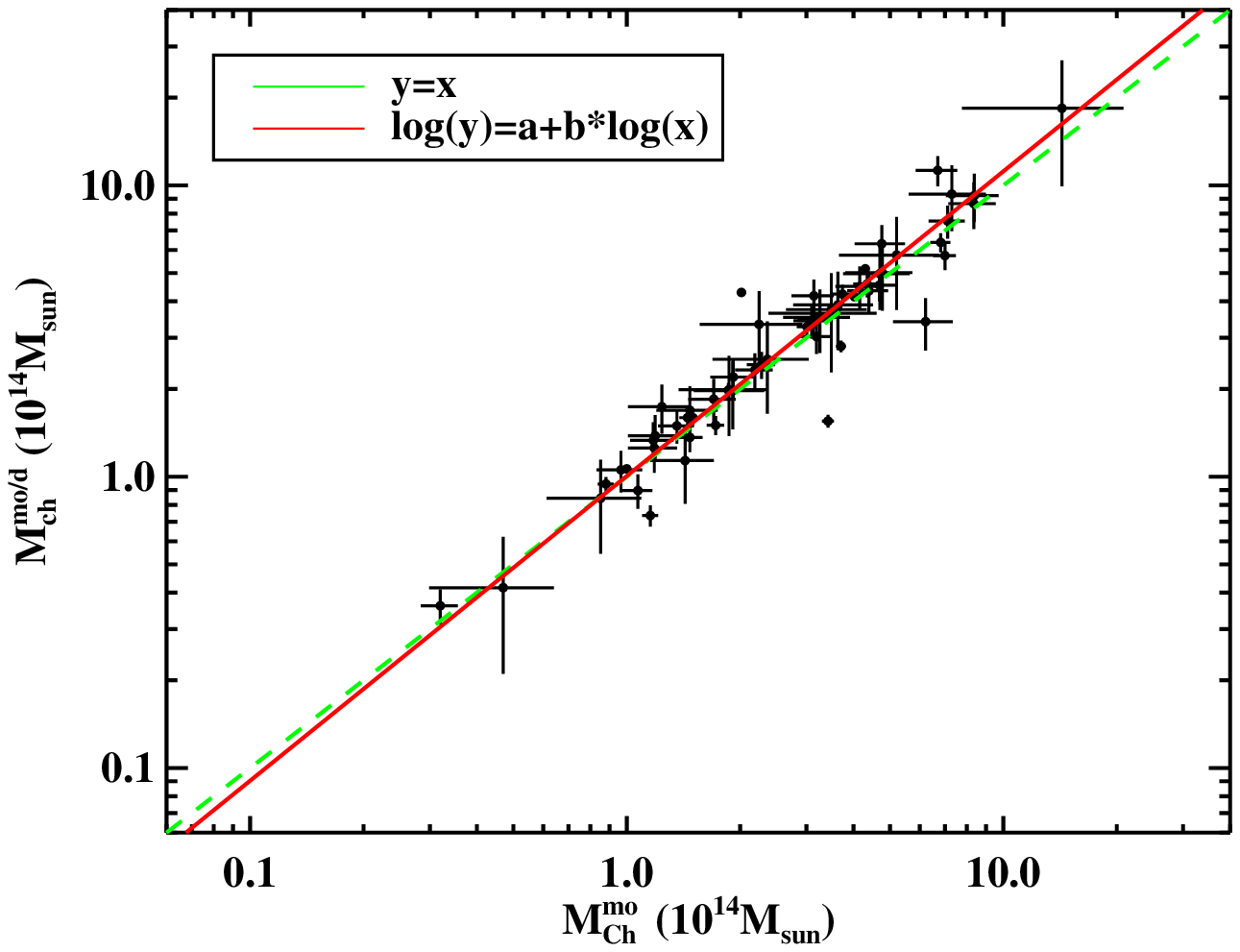,angle=0,scale=0.55}
\psfig{figure=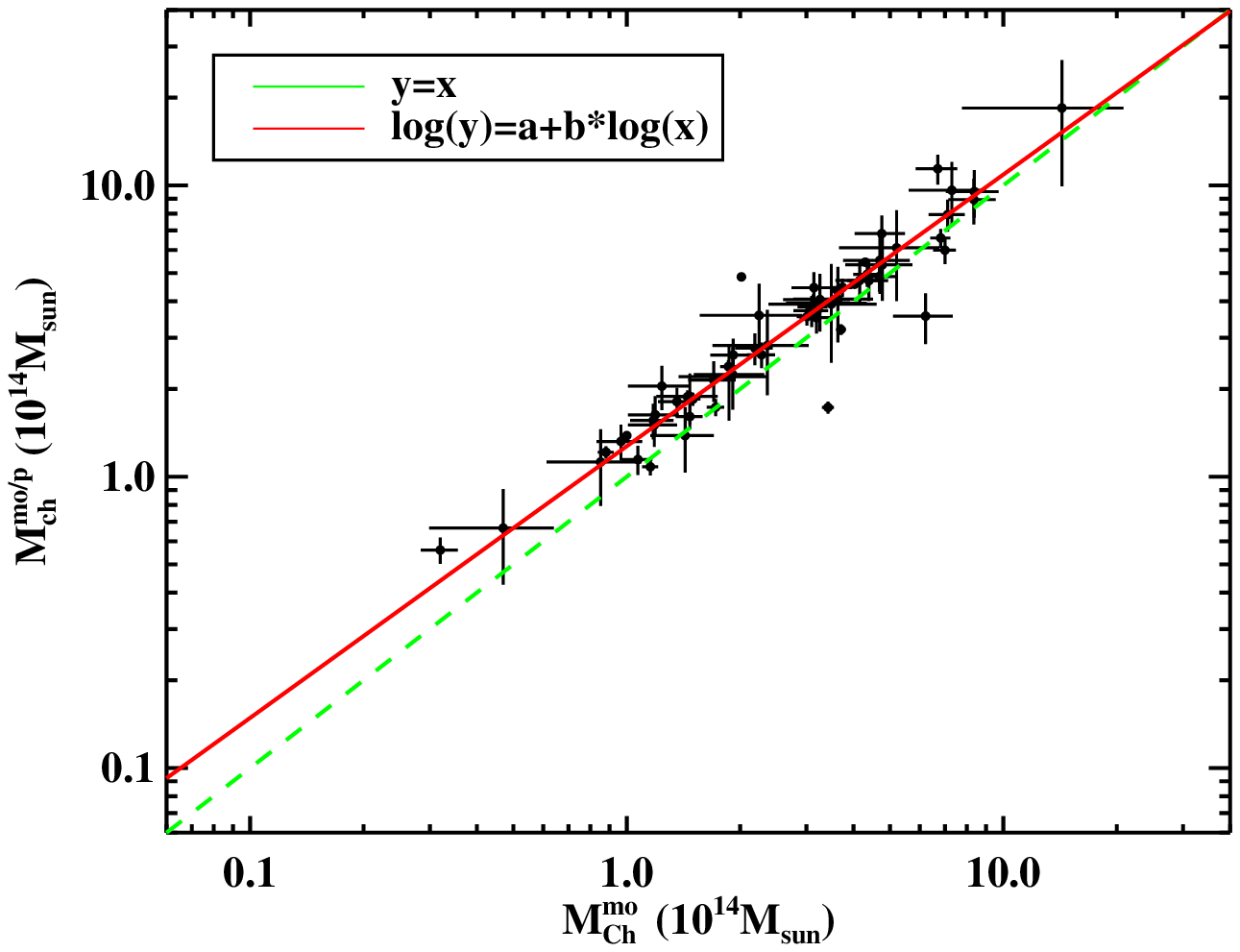,angle=0,scale=0.55}}}
\caption[Comparisons of modified {\it Chandara} masses and {\it XMM-Newton} masses.] {
Comparisons of {\it Chandara} masses modified by temperature relations and  by  our mass relation in Section 3.2.}
\label{zhaohh:fig5}
\end{figure}

We have  obtained the {\it Chandra} masses, $M_{Ch}^{mo/d}$, modified by the relation of de-projected temperatures derived
from {\it Chandra} and {\it XMM-Newton}.
In the calculating process of  $M_{Ch}^{mo/d}$, we need to modify  the {\it Chandra} temperature
and the {\it Chandra} electron density based on the modified temperature.
In addition, we can directly  get the  corresponding {\it XMM-Newton} mass (expressed as $M_{Ch}^{mo}$)
from our mass relation in Section 3.2 at given {\it Chandra} mass.
Both the methods can give the corrected {\it Chandra} masses and
we compare them in  Fig. \ref{zhaohh:fig5} to find
whether the {\it Chandra} masses  modified by  the two methods are consistent.
 The result shows that $M_{Ch}^{mo}$ is in good agreement  with  $M_{Ch}^{mo/d}$ and  the scatter  around the best-fit relation is 0.09.
It  means  that the {\it Chandra} mass corrected by our mass relation directly
is consistent with  that corrected by the complex  de-projected temperature calculation.
Thus, our mass relation is robust to correct the {\it Chandra} mass and  the relations of  de-projected temperature and mass
can provide unbiased correction for the cluster properties measured by  the two instrument.
Moreover,  based on the correction of the de-projected temperature, we can obtain  some detailed information for  clusters,
e.g., the distribution of mass, the gas fraction and entropy.

\clearpage

\section{Conclusion}

We use  a  sample of 62  galaxy clusters  to study  the discrepancies of temperature  and mass within $r_{500}$
between {\it XMM-Newton} and {\it Chandra} data.
Using the same analysis procedure, we obtain gas temperatures and total masses for each cluster with two instruments.
Comparisons of gas temperature and mass show  that:
(1) The temperatures derived from {\it Chandra} are higher than those of {\it XMM-Newton}
and there is a good linear relation for the de-projected temperature relation: $ T_{Chandra}$=1.25$\times  T_{XMM}$-0.13.
(2) The {\it Chandra} mass overestimates 36\% than the value of  {\it XMM-Newton}  and
the  relation  is $\log_{10} M_{Chandra}$=1.02 $\times \log_{10} M_{XMM}$+0.15.

To look for the possible causes of mass discrepancy, we recalculate the {\it Chandra} mass
with the de-projected and projected temperature relation (expressed as $M_{Ch}^{mo/d}$ and $M_{Ch}^{mo/p}$ ), respectively.
The results reveal that  $M_{Ch}^{mo/d}$ is more close to the $M_{XMM}$ with the bias of only 0.02.
Moreover,  the {\it Chandra} masses are recalculated with the $r_{500}$ measured by {\it XMM-Newton} and the intrinsic scatter is
significantly improved with the value reducing from  0.20 to 0.12.
Thus, the mass discrepancy is almost resolved by the temperature correction, and the temperature bias may be the main factor
causing the mass bias.
At last, we find that $M_{Ch}^{mo/d}$ is consistent with the corrected {\it Chandra} mass which is directly modified by our mass relation.
So, the de-projected temperature and original mass relation can give the unbiased corrections for galaxy cluster properties
derived from {\it Chandra} and {\it XMM-Newton}.
These relations are robust to combine {\it Chandra} and {\it XMM-Newton}  data into a large unbiased cluster sample.

This research was supported by the National Natural Science Foundation of China under
grants Nos. 11103022, and by the Strategic Priority Research Program on Space Science,
the Chinese Academy of Sciences, grant No. XDA04010300.

\begin{table} { \begin{center} \footnotesize
      {\renewcommand{\arraystretch}{1.3} \caption[]{observations and cluster properties. }
       }
\begin{tabular}{lccccccc}
\\
\hline
\hline
Name	&		redshfit	&	$N_H$	&$	T_{Chandra}			$&$	M_{Chandra}			$&$	T_{XMM}	$&$			M_{XMM}$&$R_{max}$\\
        &             & ($10^{20}{\rm cm}^{-2})$&(keV) &$(10^{14}M_{\odot})$              &(keV)         &   $(10^{14}M_{\odot})$& $(r_{500})$\\	
\hline        							
A119		&	0.044	&	3.1	&$	6.14 	\pm	0.56 	$&$	5.29 	\pm	0.16 	$&$	4.25 	\pm	0.73 	$&$	1.26 	\pm	0.17 	$&0.69	\\
A133		&	0.057	&	1.6	&$	4.89 	\pm	0.50 	$&$	2.70 	\pm	0.58 	$&$	4.04 	\pm	0.42 	$&$	3.67 	\pm	1.00 	$&0.68	\\
A1413		&	0.143	&	1.6	&$	8.39 	\pm	0.77 	$&$	5.34 	\pm	0.33 	$&$	7.01 	\pm	0.43 	$&$	6.94 	\pm	0.71 	$&1.29	\\
A1644		&	0.047	&	5.3	&$	5.00 	\pm	0.68 	$&$	1.67 	\pm	0.26 	$&$	4.50 	\pm	0.92 	$&$	2.09 	\pm	0.59 	$&0.94	\\
A1650		&	0.085	&	1.5	&$	5.90 	\pm	0.30 	$&$	4.41 	\pm	0.47 	$&$	5.49 	\pm	0.37 	$&$	3.25 	\pm	0.13 	$&1.17	\\
A1651		&	0.085	&	1.7	&$	6.48 	\pm	0.65 	$&$	6.73 	\pm	1.37 	$&$	5.42 	\pm	0.31 	$&$	3.22 	\pm	0.18 	$&0.94	\\
A1689		&	0.183	&	0.18	&$	10.94 	\pm	1.49 	$&$	10.25 	\pm	1.14 	$&$	8.69 	\pm	0.43 	$&$	8.87 	\pm	1.56 	$&1.48	\\
A1750		&	0.085	&	2.5	&$	4.65 	\pm	0.67 	$&$	2.07 	\pm	0.39 	$&$	4.18 	\pm	0.44 	$&$	0.93 	\pm	0.22 	$&1.86	\\
A1795		&	0.062	&	1.2	&$	6.32 	\pm	0.40 	$&$	6.29 	\pm	0.79 	$&$	6.75 	\pm	0.23 	$&$	7.70 	\pm	0.18 	$&0.57	\\
A1835		&	0.252	&	2.2	&$	10.66 	\pm	0.74 	$&$	10.08 	\pm	0.71 	$&$	7.58 	\pm	0.50 	$&$	6.64 	\pm	1.28 	$&2.14	\\
A1914		&	0.171	&	1	&$	10.78 	\pm	1.03 	$&$	6.81 	\pm	1.05 	$&$	8.03 	\pm	0.65 	$&$	4.52 	\pm	0.59 	$&1.74	\\
A2029		&	0.077	&	3.2	&$	8.34 	\pm	0.46 	$&$	4.47 	\pm	0.58 	$&$	6.71 	\pm	0.48 	$&$	3.08 	\pm	0.40 	$&1.13	\\
A2034		&	0.113	&	1.5	&$	7.55 	\pm	0.30 	$&$	2.85 	\pm	0.08 	$&$	6.49 	\pm	0.53 	$&$	2.86 	\pm	0.39 	$&1.40	\\
A2052		&	0.035	&	2.9	&$	3.25 	\pm	0.09 	$&$	1.62 	\pm	0.08 	$&$	3.16 	\pm	0.23 	$&$	0.86 	\pm	0.10 	$&0.76	\\
A2063		&	0.036	&	2.9	&$	5.20 	\pm	0.71 	$&$	2.04 	\pm	0.10 	 $&$	3.00 	\pm	0.78 	$&$	0.87 	\pm	0.08 	$&0.83	\\
A2065		&	0.072	&	2.8	&$	5.48 	\pm	0.67 	$&$	3.35 	\pm	0.97 	$&$	5.02 	\pm	0.34 	$&$	2.29 	\pm	0.12 	$&0.66	\\
A2142		&	0.089	&	4.1	&$	9.59 	\pm	0.59 	$&$	10.52 	\pm	2.47 	$&$	7.05 	\pm	1.11 	$&$	4.12 	\pm	0.62 	$&0.99	\\
A2147		&	0.035	&	3.29	&$	4.05 	\pm	0.48 	$&$	1.66 	\pm	0.25 	$&$	2.65 	\pm	1.11 	$&$	1.31 	\pm	0.65 	$&0.82	\\
A2163		&	0.203	&	12.3	&$	16.70 	\pm	1.22 	$&$	20.83 	\pm	9.67 	$&$	14.49 	\pm	2.02 	$&$	11.45 	\pm	3.22 	$&0.90	\\
A2199		&	0.03	&	0.8	&$	4.87 	\pm	0.22 	$&$	2.07 	\pm	0.17 	$&$	3.94 	\pm	0.20 	$&$	1.64 	\pm	0.06 	$&0.69	\\
A2204		&	0.151	&	5.9	&$	10.11 	\pm	1.44 	$&$	5.94 	\pm	0.83 	$&$	8.76 	\pm	1.23 	$&$	8.32 	\pm	0.66 	$&1.28	\\
A2219		&	0.228	&	1.7	&$	11.62 	\pm	1.64 	$&$	9.82 	\pm	0.61 	$&$	9.19 	\pm	0.79 	$&$	7.79 	\pm	0.98 	$&2.37	\\
A2255		&	0.081	&	2.5	&$	6.79 	\pm	0.24 	$&$	6.15 	\pm	0.14 	$&$	5.69 	\pm	1.20 	$&$	3.01 	\pm	0.84 	$&1.39	\\
A2256		&	0.058	&	4	&$	5.80 	\pm	0.77 	$&$	4.54 	\pm	0.45 	$&$	3.80 	\pm	0.34 	$&$	2.38 	\pm	0.25 	$&0.88	\\
A2319		&	0.056	&	8.77	&$	9.91 	\pm	1.14 	$&$	6.22 	\pm	0.51 	$&$	9.13 	\pm	0.31 	$&$	8.20 	\pm	0.14 	$&0.70	\\
A2390		&	0.233	&	10.7	&$	13.91 	\pm	0.60 	$&$	9.64 	\pm	1.24 	$&$	11.72 	\pm	1.63 	$&$	10.72 	\pm	1.32 	$&1.70	\\
A2589		&	0.042	&	4.39	&$	3.77 	\pm	0.21 	$&$	1.35 	\pm	0.19 	$&$	3.22 	\pm	0.16 	$&$	1.46 	\pm	0.13 	$&0.69	\\
A2597		&	0.085	&	2.5	&$	3.95 	\pm	0.16 	$&$	2.71 	\pm	0.36 	$&$	2.69 	\pm	0.19 	$&$	1.15 	\pm	0.07 	$&1.46	\\
A2626		&	0.057	&	92	&$	3.16 	\pm	0.20 	$&$	1.40 	 \pm	0.03 	$&$	3.32 	\pm	0.14 	$&$	1.81 	\pm	0.11 	$&0.85	\\
A2657		&	0.04	&	5.3	&$	4.21 	\pm	0.56 	$&$	2.11 	\pm	0.10 	 $&$	3.86 	\pm	0.62 	$&$	1.61 	\pm	0.32 	$&0.76	\\
A3112		&	0.075	&	2.5	&$	5.44 	\pm	0.44 	$&$	4.64 	\pm	0.95 	$&$	4.30 	\pm	0.21 	$&$	3.18 	\pm	0.17 	$&1.10	\\
A3128		&	0.062	&	1.5	&$	3.70 	\pm	0.56 	$&$	3.10 	 \pm	0.36 	$&$	2.92 	\pm	0.37 	$&$	0.90 	\pm	0.27 	$&1.30	\\
A3158		&	0.059	&	1.1	&$	5.66 	\pm	0.37 	$&$	5.75 	\pm	0.15 	$&$	4.40 	\pm	0.26 	$&$	1.94 	\pm	0.18 	$&1.03	\\
A3266		&	0.059	&	1.5	&$	10.00 	\pm	1.19 	$&$	8.94 	\pm	1.64 	$&$	7.47 	\pm	1.07 	$&$	7.31 	\pm	1.07 	$&0.76	\\
A3391		&	0.051	&	5.4	&$	6.19 	\pm	0.29 	$&$	4.88 	\pm	0.18 	$&$	5.44 	\pm	0.74 	$&$	3.74 	\pm	0.45 	$&0.95	\\
A3528		&	0.054	&	6.2	&$	5.29 	\pm	0.99 	$&$	6.83 	\pm	1.40 	 $&$	4.89 	\pm	0.71 	$&$	1.61 	\pm	0.30 	    $&1.00	\\
A3532		&	0.055	&	6	&$	5.40 	\pm	0.75 	$&$	2.40 	\pm	0.35 	$&$	4.66 	\pm	0.53 	$&$	1.20 	\pm	0.27 	$&1.31	\\
A3558B		&	0.048	&	3.6	&$	5.78 	\pm	0.45 	$&$	5.17 	\pm	0.14 	$&$	5.04 	\pm	0.59 	$&$	1.70 	\pm	0.31 	$&0.61	\\
A3562		&	0.049	&	3.9	&$	4.74 	\pm	0.51 	$&$	2.43 	\pm	0.13 	$&$	4.04 	\pm	0.21 	$&$	1.76 	\pm	0.10 	$&1.03	\\

\hline
\hline
  \end{tabular}
  \end{center}
\hspace*{0.3cm}}
\end{table}

\begin{table*} { {Table 1. continued.} \begin{center} \footnotesize
      {\renewcommand{\arraystretch}{1.3}
       }
\begin{tabular}{lccccccc}
\\
\hline
\hline
Name	&		redshfit	&	$N_H$	&$	T_{Chandra}			$&$	M_{Chandra}			$&$	T_{XMM}	$&$			M_{XMM}$&$R_{max}$\\
        &             & ($10^{20}{\rm cm}^{-2})$&(keV) &$(10^{14}M_{\odot})$              &(keV)         &   $(10^{14}M_{\odot})$& $(r_{500})$\\	
\hline
A3571		&	0.039	&	3.9	&$	8.55 	\pm	0.56 	$&$	5.19 	\pm	1.26 	$&$	5.64 	\pm	0.55 	$&$	3.25 	\pm	0.32 	$&0.58	\\
A3581		&	0.023	&	4.3	&$	2.00 	\pm	0.13 	$&$	0.44 	\pm	0.05 	$&$	2.12 	\pm	0.24 	$&$	0.79 	\pm	0.06 	$&0.56	\\
A3667		&	0.056	&	4.6	 &$	6.90 	\pm	0.35 	$&$	4.59 	\pm	0.10 	$&$	4.93 	\pm	0.26 	$&$	4.75 	\pm	0.22 	$&0.83	\\
A3827		&	0.098	&	2.8	&$	7.89 	\pm	0.57 	$&$	4.98 	\pm	1.61 	$&$	6.25 	\pm	0.48 	$&$	4.30 	\pm	0.28 	$&1.10	\\
A399		&	0.072	&	10.6	&$	8.26 	\pm	0.86 	$&$	4.28 	\pm	0.26 	$&$	5.15 	\pm	0.96 	$&$	3.05	\pm	0.64 	$&0.92	\\
%\textbf{A401}		&	0.074	&	10.6	&$	8.26 	\pm	0.86 	$&$	4.28 	\pm	0.26 	$&$	6.86	\pm	0.66 	$&$	6.30	\pm	0.35	$&0.87	\\
A4059		&	0.048	&	3.27	&$	4.88 	\pm	0.27 	$&$	1.65 	\pm	0.22 	$&$	3.96 	\pm	0.25 	$&$	1.64 	\pm	0.14 	$&0.81	\\
A478		&	0.088	&	15.3	&$	6.32 	\pm	0.26 	$&$	4.45 	\pm	0.41 	$&$	7.20 	\pm	0.61 	$&$	4.19 	\pm	0.49 	$&0.98	\\
A496		&	0.033	&	5.7	&$	5.68 	\pm	0.43 	$&$	3.18 	\pm	0.98 	$&$	3.87 	\pm	0.20 	$&$	2.11 	\pm	0.19 	$&0.96	\\
A576		&	0.038	&	5.7	&$	4.84 	\pm	0.47 	$&$	2.63 	\pm	0.13 	$&$	3.72 	\pm	0.35 	$&$	1.06 	\pm	0.10 	$&0.96	\\
A754		&	0.054	&	4.6	&$	11.58 	\pm	1.37 	$&$	12.06 	\pm	1.76 	$&$	8.90 	\pm	0.93 	$&$	6.97 	\pm	2.59 	$&0.57	\\
A85		&	0.06	&	3.6	&$	6.14 	\pm	0.32 	$&$	3.23 	\pm	0.28 	$&$	5.28 	\pm	0.22 	$&$	2.63 	\pm	0.09 	$&0.70	\\
EXO0422		&	0.039	&	6.4	&$	3.86 	\pm	0.33 	$&$	2.01 	\pm	0.39 	$&$	3.01 	\pm	0.18 	$&$	1.28 	\pm	0.12 	$&0.69	\\
HERCULES		&	0.037	&	3.4	&$	2.88 	\pm	0.80 	$&$	0.65 	\pm	0.24 	$&$	2.71 	\pm	0.78 	$&$	0.88 	\pm	0.29 	$&0.73	\\
HydraA		&	0.054	&	4.9	&$	3.80 	\pm	0.11 	$&$	1.74 	\pm	0.33 	$&$	3.52 	\pm	0.11 	$&$	1.85 	\pm	0.08 	$&0.81	\\
IIIZW54		&	0.031	&	16.68	&$	3.04 	\pm	0.28 	$&$	1.19 	\pm	0.34 	$&$	2.36 	\pm	0.18 	$&$	1.08 	\pm	0.14 	$&0.68	\\
MKW3s		&	0.044	&	3.2	&$	4.12 	\pm	0.18 	$&$	1.91 	\pm	0.21 	$&$	3.48 	\pm	0.15 	$&$	1.80 	\pm	0.35 	$&0.68	\\
MKW8		&	0.026	&	2.6	&$	3.28 	\pm	0.14 	$&$	1.23 	\pm	0.06 	$&$	2.95 	\pm	0.27 	$&$	0.67 	\pm	0.08 	$&0.71	\\
PKS0745		&	0.103	&	43.49	&$	8.77 	\pm	0.79 	$&$	6.71 	\pm	0.69 	$&$	6.48 	\pm	0.43 	$&$	3.61 	\pm	0.25 	$&1.19	\\
RXCJ1504.1-0248		&	0.215	&	6	&$	11.64 	\pm	0.95 	$&$	12.10 	\pm	1.96 	$&$	7.36 	\pm	0.41 	$&$	5.07 	\pm	0.66 	$&2.04	\\
RXCJ1558.3-1410		&	0.097	&	11.1	&$	5.23 	\pm	0.27 	$&$	2.64 	\pm	0.71 	$&$	4.36 	\pm	0.51 	$&$	1.72 	\pm	0.16 	$&1.71	\\
RXCJ1720.1+2637		&	0.164	&	3.9	&$	8.26 	\pm	1.59 	$&$	7.47 	\pm	2.25 	$&$	7.92 	\pm	0.71 	$&$	5.24 	\pm	0.57 	$&1.60	\\
RXCJ2014.8-2430		&	0.16	&	13.22	&$	7.37 	\pm	0.57 	$&$	4.98 	\pm	1.22 	$&$	5.66 	\pm	0.35 	$&$	4.19 	 \pm	0.70 	$&1.70	\\
S1101		&	0.056	&	1.9	&$	2.89 	\pm	0.11 	$&$	1.50 	 \pm	0.14 	$&$	2.36 	\pm	0.06 	$&$	1.51 	\pm	0.16 	$&0.91	\\
\hline
2A0335		&	0.035	&	18.6	&$	3.86 	\pm	0.25 	$&$	1.43 	\pm	0.04 	$&$	4.03 	\pm	0.80 	$&$	2.96 	\pm	0.44 	$&0.45	\\
A1060		&	0.013	&	4.9	&$	4.33 	\pm	1.25 	$&$	1.24 	\pm	0.28 	$&$	3.15 	\pm	0.25 	$&$	0.93 	\pm	0.06 	$&0.34	\\
A262		&	0.016	&	5.5	&$	2.84 	\pm	0.12 	$&$	0.80 	\pm	0.05 	$&$	2.48 	\pm	0.33 	$&$	0.58 	\pm	0.07 	$&0.37	\\
A3526		&	0.011	&	8.2	&$	3.09 	\pm	0.16 	$&$	0.82 	\pm	0.03 	$&$	3.19 	\pm	0.21 	$&$	1.10 	\pm	0.06 	$&0.29	\\
AWM7		&	0.017	&	9.21	&$	4.25 	\pm	0.39 	$&$	1.42 	\pm	0.19 	$&$	3.59 	\pm	0.57 	$&$	1.31 	\pm	0.15 	$&0.36	\\
COMA		&	0.023	&	0.9	&$	8.97 	\pm	0.19 	$&$	8.45 	\pm	0.30 	 $&$	7.14 	\pm	0.47 	$&$	8.89 	\pm	1.15 	$&0.29	\\
MKW4		&	0.02	&	1.9	&$	1.81 	\pm	0.10 	$&$	0.56 	\pm	0.07 	$&$	2.04 	\pm	0.29 	$&$	0.62 	\pm	0.12 	$&0.44	\\
NGC1550		&	0.013	&	11.6	&$	1.68 	\pm	0.25 	$&$	0.55 	\pm	0.09 	$&$	1.44 	\pm	0.06 	$&$	0.25 	\pm	0.01 	$&0.47	\\
NGC5813   		&	0.007	&	4.2	&$	0.72 	\pm	0.14 	$&$	0.11 	\pm	0.01 	$&$	0.75 	\pm	0.01 	$&$	0.17 	\pm	0.02 	$&0.23	\\
OPHIUCHUS 		&	0.028	&	20.14	&$	8.42 	\pm	0.17 	$&$	3.97 	\pm	0.18 	$&$	8.65 	\pm	0.38 	$&$	12.50 	\pm	0.30 	$&0.45	\\

\hline
\hline
\end{tabular}
\end{center}
\hspace*{0.3cm}{\footnotesize \textbf{Notes.} The redshfits of  clusters used in this sample are quoted from NASA/IPAC Extragalactic Database (NED).
   The hydrogen absorption columns, $N_H$, are obtained from Dickey \& Lockman (1990). $r_{500}$ are measured by {\it XMM-Newton}.
   $T_{Chandra}$ and $T_{XMM}	$ are the global temperatures defined in Section 3.1, $	M_{Chandra}$  and $M_{XMM}$ represent the original
  {\it Chandra} and {\it XMM-Newton} masses within  their own $r_{500}$. $R_{max}$ is the maximum radius, out to which the temperature profile can reach
 for each cluster as a fraction of $r_{500}$. Ten clusters, which $R_{max}$ are smaller than 0.5$r_{500}$, are listed in the end of table.}}
\end{table*}
%-------------------------------------------------

\begin{table}
%\begin{center}
\renewcommand{\arraystretch}{1.05}
\caption[]{Comparisons of parameters derived from {\it Chandra} and {\it XMM-Newton}.}
\begin{tabular}{lcccc}
\\
\hline
\hline
        &A & B    & intrinsic scatter & bias\\
\hline
$\emph{\textbf{T}}_{\emph{\textbf{XMM}}}$-$\emph{\textbf{T}}_{\emph{\textbf{Chandra}}}$ (dep) &$-0.13\pm 0.26$ & 1.25$\pm$0.07 & $0.50\pm 0.02$ & $1.24\pm 0.07$ \\
$\emph{T}_{\emph{XMM}}$-$\emph{T}_{\emph{Chandra}}$ (p)      &$-0.83\pm 0.32$ & $ 1.30 \pm 0.08 $  & $0.57\pm 0.01$ & $0.79\pm 0.05$ \\
\hline
$\emph{\textbf{M}}_{\emph{\textbf{XMM}}}$-$\emph{\textbf{M}}_{\emph{\textbf{Chandra}}}$         &$0.15\pm 0.05$  &1.02$\pm$0.09       & $0.19\pm 0.01$ & $0.15\pm 0.01$ \\
\hline
$M_{XMM}$-$M_{Ch}^{mo/d} $               &$0.01\pm0.05$ & 1.05$\pm$0.08      & $0.20\pm 0.01$ & $0.02\pm0.01$ \\
$M_{XMM}$-$M_{Ch}^{mo/p} $           &$0.10\pm0.04$ & 0.93$\pm$0.07       & $0.18\pm 0.01$ &$0.08\pm0.01$ \\
\hline
$M_{XMM}$-$M_{Ch}^{mo/d,r} $            &$0.04\pm0.03$ & 0.97$\pm$0.06     & $0.12\pm 0.01$ &$0.03\pm0.01$ \\
$M_{XMM}$-$M_{Ch}^{mo/p,r} $            &$0.13\pm0.03$ & 0.89$\pm$0.06       & $0.11\pm 0.01$ &$0.08\pm0.01$ \\
\hline
$M_{Ch}^{mo} $-$M_{Ch}^{mo/d} $        &$0.01\pm0.02$  &1.05$\pm$0.03       & $0.09\pm 0.01$ & $0.02\pm0.01$ \\
$M_{Ch}^{mo} $-$M_{Ch}^{mo/p} $         &$0.10\pm0.02$  &0.93$\pm$0.04       & $0.09\pm 0.01$ & $0.08\pm0.01$ \\
\hline
\hline
\\
\noalign{\footnotesize  \textbf{Note.}
dep: the temperatures in the relation are de-projected temperatures. p: the temperatures  are projected temperature.
$M_{XMM}$ and $M_{Chandra}$ represent the original de-projected masses determined by {\it Chandra} and {\it XMM-Newton}, respectively.
$M_{ch}^{mo/d}$ and $M_{ch}^{mo/p}$ are the {\it Chandra} masses modified by the de-projected and projected temperature relations, respectively.
$M_{ch}^{mo/d,r}$: the $M_{ch}^{mo/d}$ is recalculated  with the $r_{500}$ measured by {\it XMM-Newton}.
$M_{ch}^{mo/p,r}$: the $M_{ch}^{mo/p}$ is recalculated  with the $r_{500}$ measured by {\it XMM-Newton}.
 The $M_{Ch}^{mo} $ is derived from the  relation of $M_{XMM}$-$M_{Chandra} $ at given $M_{Chandra}$.
A linear fitting, $Y$=B$\times$$X$+A , is used for the temperature relations.
We perform the mass relations with  logarithmic values of the parameters in the form: log$_{10} Y$=B$\times$log$_{10} X$+A.
 The bias represents the systematic deviation in temperature or  mass measurement between different instruments.
 The relations which can provide unbiased corrections for the properties measured by {\it Chandra} and {\it XMM-Newton} are
highlighted with bold letters.
 }
\end{tabular}
%\end{center}
\end{table}

\end{document}